\documentstyle[aps,prl,twocolumn,epsf,floats]{revtex}

\begin{document}
\draft

\wideabs{

\title{Composite Fermion Approach to the Quantum Hall Hierarchy: \protect\\
       When it Works and Why}

\author{
   Arkadiusz W\'ojs$^*$ and John J. Quinn}
\address{
   The University of Tennessee, Knoxville, Tennessee, 37996}

\maketitle

\begin{abstract}
The mean field composite Fermion (MFCF) picture has been qualitatively
successful when applied to electrons (or holes) in the lowest Landau level.
Because the energy scales associated with Coulomb interactions and with
Chern--Simons gauge field interactions are different, there is no rigorous
justification of the qualitative success of the MFCF picture.
Here we show that what the MFCF picture does is to select from all the 
allowed angular momentum ($L$) multiplets of $N$ electrons on a sphere, 
a subset with smaller values of $L$. 
For this subset, the coefficients of fractional parentage for pair states
with small relative angular momentum $R$ (and therefore large repulsion)
either vanish or they are small.
This set of states forms the lowest energy sector of the spectrum.
\end{abstract}

}

When applied to electrons (or holes) in the lowest Landau level, the mean
field composite Fermion (MFCF) picture \cite{jain} gives a good qualitative
description of the low lying states of fractional quantum Hall systems
\cite{tsui}.
The original conjecture that the CF transformation converts a system of 
strongly interacting electrons into one of weakly interacting composite 
Fermions cannot possibly be correct because the Chern--Simons interactions 
among fluctuations are measured on an energy scale (proportional to magnetic
field $B$) which can be much larger than the energy scale of the Coulomb 
interactions (proportional to $B^{1/2}$).
Because so many results, both of large numerical calculations and of
experiments, can be interpreted in terms of composite Fermions, it is
extremely important to understand why the MFCF picture works.
This is the problem we address in this letter.

For $N$ electrons on a Haldane sphere \cite{haldane} (containing at the
center a magnetic monopole of charge $2S\,hc/e$), the single particle
states fall into angular momentum shells with $l_n=S+n$, $n=0,1,\dots$
The $n$th shell is $2l_n+1$ fold degenerate.
The CF transformation attaches to each electron a flux tube of strength 
$2p_0$ flux quanta oriented opposite to the original magnetic field.
If the added flux is treated in a mean field approximation, the resulting
effective magnetic field seen by one CF is $B^*=B-2p_0\,(hc/e)\,n_s$
($n_s$ is the number of electrons per unit area). 
An effective CF filling factor $\nu_0^*$ (${\nu_0^*}^{-1}=\nu_0^{-1}-2p_0$) 
and an effective monopole strength $2S^*$ ($2S^*=2S-2p_0(N-1)$) seen by 
one CF can also be defined.
$|S^*|$ plays the role of the angular momentum of the lowest CF shell
\cite{chen}.
States belonging to the Jain sequence occur when $\nu_0^*$ is an integer.
For such integral CF fillings, the ground state is a Laughlin 
\cite{laughlin} incompressible liquid state with angular momentum $L=0$.
If $\nu_0^*$ is not an integer, a partially occupied CF shell will contain 
$n_{\rm QE}$ quasielectrons (or $n_{\rm QH}$ quasiholes).
In the MFCF picture these states form a degenerate band of angular 
momentum multiplets with energy $n_{\rm QE}\varepsilon_{\rm QE}$ where 
$\varepsilon_{\rm QE}$ is the energy of a single quasielectron (or
$n_{\rm QH}\varepsilon_{\rm QH}$ for quasiholes).
The degeneracy results from the neglect of QP--QP interactions in the MFCF
approximation \cite{sitko1}.

Hierarchy states \cite{sitko2} outside the Jain sequence are obtained
(when $\nu_0^*$ is not equal to an integer) by reapplying the CF
transformation to residual quasiparticles in the partially filled CF shell.
In comparing the predictions of the CF hierarchy picture with numerical
results for states containing three or four quasiparticles, it is found
that the MF approximation is often qualitatively incorrect.
Before worrying about the reapplication of the MFCF approach to residual
quasiparticles in a partially filled CF shell, it is very useful to ask why 
the MFCF picture applied directly to electrons (or holes) in a partially
filled shell gives qualitatively correct results.
In light of the different energy scales describing Coulomb interactions
and Chern--Simons gauge field interactions, the justification cannot lie
in a cancellation between these interactions.

\begin{figure}[t]
\epsfxsize=3.25in
\epsffile{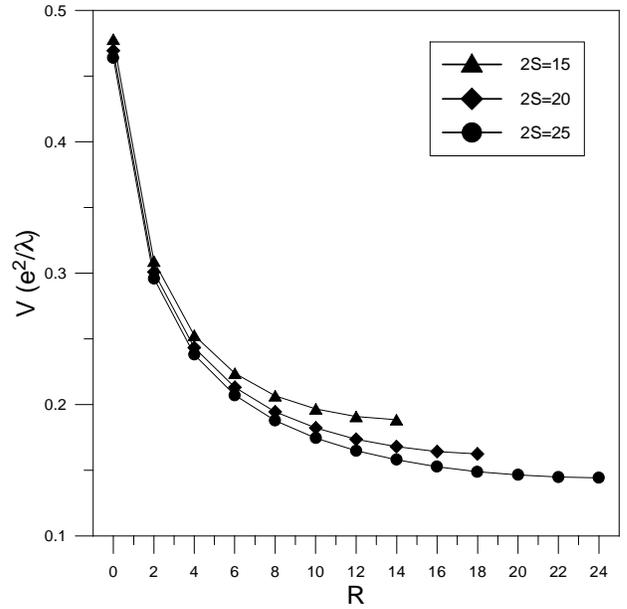}
\caption{
   Plot of $V(R)$, the pseudopotential coefficient of the Coulomb 
   interaction potential, as a function of relative angular momentum
   $R=L^{\rm MAX}-L$ of a pair of electrons, for $2S=15$, 20, and 25.
   $L$ is the pair angular momentum and $L^{\rm MAX}=2l-1$ its maximum 
   value.
   $V(R)$ is measured in units of $e^2/\lambda$, where $\lambda$ is the
   magnetic length}
\label{fig1}
\end{figure}

We begin by considering the simple case of three electrons in the lowest
angular momentum shell, with values of $2S$ ranging from 2 to 14. 
In Tab.~\ref{tab1} we give the number of times an $L$-multiplet occurs for 
a given value of $2S$ (upper table is for even values of $2S$ and lower one 
for odd values).
Note that the set of values of $L$ at $2S-2(N-1)=2S-4$ is always a subset
of the set at $2S$.
\begin{table}
\caption{
   The number of times an $L$-multiplet appears for a system of three
   electrons of angular momentum $l=S$.
   Top: even values of $2S$; bottom: odd values of $2S$.
   Blank spaces are equivalent to zeros}
\begin{tabular}{r|ccccccccccccccccccc}
  $_{2S}\mbox{}^{2L}$
     & 0& 2& 4& 6& 8&10&12&14&16&18&20&22&24&26&28&30&32&34&36\\ \hline
    2& 1&  &  &  &  &  &  &  &  &  &  &  &  &  &  &  &  &  &  \\ 
    4&  & 1&  & 1&  &  &  &  &  &  &  &  &  &  &  &  &  &  &  \\ 
    6& 1&  & 1& 1& 1&  & 1&  &  &  &  &  &  &  &  &  &  &  &  \\ 
    8&  & 1&  & 2& 1& 1& 1& 1&  & 1&  &  &  &  &  &  &  &  &  \\ 
   10& 1&  & 1& 1& 2& 1& 2& 1& 1& 1& 1&  & 1&  &  &  &  &  &  \\ 
   12&  & 1&  & 2& 1& 2& 2& 2& 1& 2& 1& 1& 1& 1&  & 1&  &  &  \\ 
   14& 1&  & 1& 1& 2& 1& 3& 2& 2& 2& 2& 1& 2& 1& 1& 1& 1&  & 1 
\end{tabular}
\begin{tabular}{r|ccccccccccccccccccc}
  $_{2S}\mbox{}^{2L}$
     & 1& 3& 5& 7& 9&11&13&15&17&19&21&23&25&27&29&31&33&35&37\\ \hline
    3&  & 1&  &  &  &  &  &  &  &  &  &  &  &  &  &  &  &  &  \\ 
    5&  & 1& 1&  & 1&  &  &  &  &  &  &  &  &  &  &  &  &  &  \\ 
    7&  & 1& 1& 1& 1& 1&  & 1&  &  &  &  &  &  &  &  &  &  &  \\ 
    9&  & 1& 1& 1& 2& 1& 1& 1& 1&  & 1&  &  &  &  &  &  &  &  \\ 
   11&  & 1& 1& 1& 2& 2& 1& 2& 1& 1& 1& 1&  & 1&  &  &  &  &  \\ 
   13&  & 1& 1& 1& 2& 2& 2& 2& 2& 1& 2& 1& 1& 1& 1&  & 1&  &  
\end{tabular}
\label{tab1}
\end{table}
The validity of this result can be easily established numerically
for arbitrary values of $N$ and $2S$ (including values of $N$ larger
than $4(S-p_0(N-1)+1)$, where more than one CF shell is filled).
In Fig.~\ref{fig1} the pseudopotential coefficient $V(R)$ is plotted for 
$2S=15$, 20, and 25, as a function of $R$, the relative angular momentum 
of a pair of electrons.
$R$ is defined as \cite{haldane,sitko1} $L^{\rm MAX}-L$, where $L^{\rm 
MAX}$ is the maximum possible angular momentum of a pair of electrons
each with angular momentum $l$, and $L$ is the total angular momentum of
the pair.
$R$ takes on even values less than or equal to $2l-1$.
$V(R)$ is a monotonically decreasing function of $R$ (for the lowest
angular momentum shell), and $V(0)$ is considerably larger that $V(2)$.
In low-lying energy states, electrons have to avoid having $R=0$ because
of the large repulsion.

An antisymmetric eigenfunction $\left|l^3,L\alpha\right>$ of three
electrons each of angular momentum $l$ whose total angular momentum is 
$L$ can be written as \cite{shalit}
\begin{equation}
\label{eq1}
   \left|l^3,L\alpha\right>=\sum_{L'}F_{L\alpha}(L')
   \left|l^2,L';l,L\right>.
\end{equation}
Here $\alpha$ is an index which distinguishes different multiplets with 
the same total angular momentum $L$.
$\left|l^2,L';l,L\right>$ is the state in which $l_1=l_2=l$ are added to
obtain pair angular momentum $L'$, and then $l_3=l$ is added to $L'$
to obtain angular momentum $L$.
Note that even though $\left|l^2,L';l,L\right>$ is not antisymmetric 
because $l_3$ is treated differently from $l_1$ and $l_2$ ($\left|l^2,
L'\right>$ is antisymmetric under interchange of 1 and 2), $\left|l^3,
L\alpha\right>$ is antisymmetric.
The factor $F_{L\alpha}(L')$ is called the coefficient of fractional 
parentage (CFP) associated with pair angular momentum $L'$. 
The two-particle interaction matrix element can be conveniently expressed
via the CFP's and the pseudopotential coefficients \cite{sitko1},
\begin{equation}
\label{eq2}
   \left<l^3,L\alpha\right|V\left|l^3,L\beta\right>
   =3\sum_{L'} F_{L\alpha}(L')F_{L\beta}(L')\,V(R).
\end{equation}
Because of the large Coulomb repulsion at $R=0$, the low lying states will
avoid pair angular momentum $L'=L'^{\rm MAX}$ (corresponding to $R=0$).
Where can such states occur?
If we choose $L'=L'^{\rm MAX}$, then $L$, the total angular momentum, must
be greater than or equal to $L'^{\rm MAX}-l$, the minimum possible value of 
the addition of $L'^{\rm MAX}$ and $l$, the angular momentum of the third
electron.
Because $L'^{\rm MAX}=2l-1$, the minimum possible value of $L$ for which
$R=0$ is $L=l-1$.
States with $L<l-1$ must have $R\ge2$.
Note that although we have selected the pair ($l_1,l_2$) to give $L'$,
the CFP's give an eigenfunction of $L$ which is totally antisymmetric.
Therefore we need not worry about other pair angular momenta in writing
down Eq.~(\ref{eq1}).
The next lower value of $L'$ is $L'^{\rm MAX}-2$ (corresponding to $R=2$)
and states with $L<l-3$ must have $R\ge4$.
Further, states with $L<l-5$ must have $R\ge6$, and so on.
In Tab.~\ref{tab2} we list the values of $2L$ for which the CFP with $R=0$
must vanish (i.e. $2L\;(R\ge2)$), for which the CFP with $R\le2$ must vanish,
and for which the CFP with $R\le4$ must vanish.
The $L=0$ state for $2S=6$ is the Laughlin $\nu=1/3$ state, for $2S=10$ it 
is the $\nu=1/5$ state, and for $2S=14$ it is the $\nu=1/7$ state.
\begin{table}[b]
\caption{
   The allowed values of $2L$ for a three electron system that must have 
   $R\ge2$, $R\ge4$, and $R\ge6$.
   The listed $2L$ values are always a subset of the allowed 
   $L$-multiplets given in Tab.~\protect\ref{tab1}}
\begin{tabular}{c|lllllllll}
  $2l=2S$   & 6& 7& 8& 9&10&11&12&13&14 \\ \hline
  $2L\;(R\ge2)$ & 0& 3& 2& 3,5& 0,4,6& 3,5,7& 2,6$^2$,8
            & 3,5,7,9$^2$& 0,4,6,8$^2$,10 \\
  $2L\;(R\ge4)$ & & & & & 0& 3& 2& 3,5& 0,4,6 \\
  $2L\;(R\ge6)$ & & & & & & & & & 0
\end{tabular}
\label{tab2}
\end{table}
At $2S=8$ two $L=3$ multiplets occur (see Tab.~\ref{tab1}).
The interparticle interaction must be diagonalized in this two-dimensional 
subspace.
We find that for the linear combination with the lower eigenvalue, the 
CFP for $R=0$ almost vanishes (its value is less than 0.001).
A similar thing occurs at $2S= 9$ for $L=9/2$, at $2S=10$ for $L=4$ and 6,
at $2S=11$ for $L=9/2$, 11/2, and 15/2, at $2S=12$ for $L=5$, 6, 7, and 9,
at $2S=13$ for $L=11/2$, 13/2, 15/2, 17/2, and 21/2, and at $2S=14$ for 
$L=6^2$, 7, 8, 9, 10, and 12.
At $2S=14$ for $L=6$ there are three allowed multiplets, and diagonalization
of the Coulomb interaction gives CFP for $R=0$ which are very small for two
of these states.
One can see that the subset of states at $2S-2(N-1)=2S-4$ of the allowed 
states at $2S$ all have CFP for $R=0$ which either vanishes identically or 
is very small (due to diagonalization of the Coulomb interaction in the
subspace of a given $L$).
But $2S-2(N-1)$ is just $2S^*$ for $2p_0=2$.
Thus the CF picture simply picks the subset of angular momentum multiplets
which has $F_{L\alpha}(L')$ essentially equal to zero for $L'=L'^{\rm MAX}$
or $R=0$.
For the $\nu=1/5$ (or $\nu=1/7$) state, the effective monopole strength 
$2S^*=2S-4(N-1)$ (or $2S^*=2S-6(N-1)$) picks the subset of states with 
$R\ge4$ (or $R\ge6$).
The MFCF picture assumes the CFP's for $R<2p_0$ to vanish and simply 
neglects $V(R)$ for larger $R$.

\begin{figure}[t]
\epsfxsize=3.25in
\epsffile{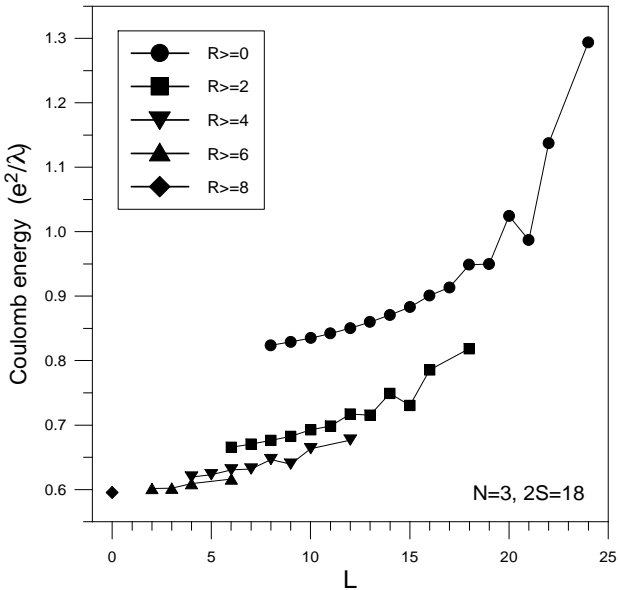}
\caption{
   The Coulomb energy of three electrons with $2S=18$.
   The highest energy states have fractional parentage with $R=0$ (dots),
   next highest with $R=2$ (squares), etc., down to the $L=0$ ground state
   (diamond) which has no parentage with $R<8$}
\label{fig2}
\end{figure}

In Fig.~\ref{fig2} we plot the Coulomb energy as a function of $L$ for
the three electron system with $2S=18$.
The dots are $L$-multiplets that have some fractional parentage at $R=0$,
the squares have some fractional parentage at $R=2$ but not at $R=0$, etc.
Clearly, the gap associated with $V(R=0)$ is the largest, that with
$V(R=2)$ the next largest, etc.
The $L=0$ ground state (marked by a diamond) corresponds to $\nu=1/9$.
Its gap is associated with $V(R=8)$ and it is almost unobservable as 
might be expected.
Note that the first excited band in this figure (states with $R\ge6$)
contains multiplets at $L=2$, 3, 4, 6, in contrast with the MFCF 
prediction ($L=1$, 2, 3).

\begin{figure}[t]
\epsfxsize=3.25in
\epsffile{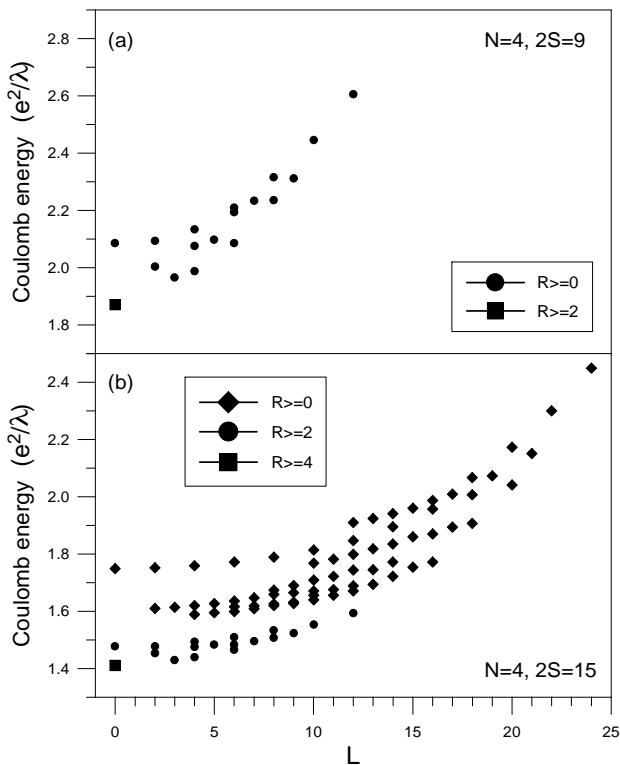}
\caption{
   Energy spectra, $E$ vs. $L$, for a four electron system with $2S=9$
   (a) and $2S=15$ (b).
   The states in (a) are a subset of those in (b).
   In (a) only the lowest state has $R\ge2$ (square).
   In (b) the lowest state has $R\ge4$ (square) and the band of states 
   marked by dots have $R\ge2$}
\label{fig3}
\end{figure}

Up to here we have concentrated on the three electron system because it 
is simple and contains the essential physics of our ideas on the validity 
of MFCF picture.
Note that we have not introduced a second energy scale (proportional to 
$B$); the only energy scale is $e^2/\lambda$ ($\lambda$ is the magnetic
length) and the Coulomb interaction is analyzed in terms of its
pseudopotential coefficients $V(R)$.
Can we go beyond the three electron system?
The answer is yes.
The coefficients of fractional grandparentage (CFGP), equivalent to 
the CFP's in the case of three particles, allow us to write the totally
antisymmetric $N$-particle wavefunction as a combination of wavefunctions 
which are antisymmetric under interchange of particles 1 and 2 and under
interchange of particles 3, 4, $\dots$ , $N$, and which have well defined
angular momentum $L'$ of the pair (1,2).
The wavefunction and energy expansions are analogous to those given in
Eqs.~(\ref{eq1}) and (\ref{eq2}),
\begin{eqnarray}
\label{eq3}
   & &\left|l^N,L\alpha\right>=
   \sum_{L'}\sum_{L''\alpha''} 
   G_{L\alpha,L''\alpha''}(L')
   \left|l^2,L';l^{N-2},L''\alpha'';L\right>,
\\
\label{eq4}
   & &\left<l^N,L\alpha\right|V\left|l^N,L\beta\right>=
   {N(N-1)\over2}\times
\nonumber\\
   & &\sum_{L'} \left( \sum_{L''\alpha''}
   G_{L\alpha,L''\alpha''}(L')\; 
   G_{L\beta, L''\alpha''}(L') \right) V(R).
\end{eqnarray}
Tables of CFP's and CFGP's are given in nuclear and atomic physics books 
\cite{shalit}.
If our picture is correct, an $N$ particle system should have bands of
states with some fractional parentage for $R\ge0$, others with $R\ge2$, etc.
In Fig.~\ref{fig3} we show the four electron spectra for $2S=9$ (Laughlin
$\nu=1/3$ state) and $2S=15$ (Laughlin $\nu=1/5$ state).
In the former case only the low energy $L=0$ state (marked by a square)
has $R\ge2$, all other states have some parentage from $R=0$.
In the latter case the entire subset of states which appeared at $2S=9$
form a low energy band with $R\ge2$, and the lowest $L=0$ state has 
$R\ge4$.

The CFP's and CFGP's can be found in books or computed directly.
When a value of $L$ contains a number of multiplets, the set of basis
functions and the CFGP's are not uniquely defined in the absence of
interaction, which must be diagonalized within the $L$-subspace.
We were able to use our numerical code to determine the squares of the
CFGP's by diagonalizing the Coulomb interaction, and then by calculating 
the matrix elements of an interaction where we set $V=1$ for a given 
$R$ and $V=0$ elsewhere.
As follows from Eq.~(\ref{eq4}), the expectation values obtained in this 
way determine the squares of CFGP's.
In all cases where we expect the CFGP to vanish or to be small compared
to the values for neighboring $L$-multiplets, we find that it is so, for 
$N$ up to 8 and for values of $\nu=1/3$, $2/3$, $2/5$, $2/7$, and $2/9$.

Two other points are worth mentioning.
First, the idea of choosing a subset of the allowed values of $L$ for $N$
Fermions each with angular momentum $l$ by looking only at states that can
be obtained by adding the angular momentum of $N$ Fermions with $l^*=l-p_0
(N-1)$ is intimately connected with avoiding fractional parentage with
$R=0$, or $R=2$ (for $\nu\sim1/5$), etc.
To have the same concept be valid for residual quasiparticles in a 
partially filled CF shell (which would generate CF hierarchy states 
\cite{sitko2} like $\nu=4/11$ or $4/13$ that are outside the Jain 
sequence), the QP--QP interaction would have to be similar to the Coulomb 
interaction (i.e. strongly repulsive for $R=0$ and decreasing in magnitude 
with increasing $R$).
This does not occur for quasielectrons in the first excited CF shell, but
does occur for quasiholes in the lowest CF shell \cite{sitko2}.
Thus it is clear why the CF hierarchy scheme is often invalid for states
outside the Jain sequence.
Second, states with a single quasihole (e.g. in the $\nu=1/3$ state) have 
$R$ greater than or equal to the value in the neighboring Laughlin state 
($R\ge2$ for the $\nu=1/3$ state), while for a single quasielectron there 
has to be some parentage for $R$ less than this value (for $R=0$ in the 
single quasielectron state adjacent to $\nu=1/3$).
This explains why the quasielectron energy is larger than the quasihole
energy; $V(R=0)$ is much larger than $V(R=2)$.

We have shown that the qualitative results of the MFCF picture can be
justified by considering the subset of $L$-multiplets obtained by adding
$N$ angular momenta $l^*=l-p_0(N-1)$.
These states have smaller total angular momentum and larger values of 
the relative angular momentum of pair states.
The $L=0$ incompressible ground states at $\nu=(1+2p_0)^{-1}$ have 
fractional parentage for values of $R<2p_0$ which is essentially equal 
to zero, in agreement with the correlation effects proposed in Laughlin's
original paper \cite{laughlin}.
In fact, for model interactions in which $V(R)$ decreases very rapidly 
with increasing $R$ (i.e. for very short range interactions), exact
diagonalization gives exact zeros of the coefficients of fractional 
parentage for $R<2p_0$ instead of the very small values obtained with
the Coulomb interaction.
For fractions containing an integer other than unity in the numerator,
we find that the CFGP of one $L=0$ state is small compared to those of
other $L$-multiplets, leading to a low-lying incompressible ground state.
Only a single energy scale, $e^2/\lambda$, the Coulomb scale, is involved
in the analysis.
The CF transformation is a convenient way to arrive at a subset of the
allowed $L$ values, however the energy scale of the non-interacting
composite Fermions is totally irrelevant.

This work was supported in part by the Division of Materials Sciences 
-- Basic Energy Research Program of the U.S. Department of Energy.
A. W. acknowledges support from the KBN Grant No. PB674/P03/96/10.
The authors would like to thank P.Sitko, K. S. Yi, and D. C. Marinescu
for discussions on the preliminary aspects of this work.

$^*$ On leave from the Institute of Physics, Wroc\l aw University of 
Technology, Wroc\l aw, Poland.

\end{document}